\documentclass{article}

\usepackage[margin=0.95in]{geometry}
\usepackage[utf8]{inputenc}
\usepackage{amsmath}
\usepackage{amssymb}
\usepackage{amsthm}
\usepackage{graphicx}
\usepackage{subcaption}
\usepackage{array,multirow}
\usepackage[numbers]{natbib}
\usepackage{float}

\usepackage[margin=1cm]{caption}

\usepackage{subcaption}
\usepackage{float}

\newcommand{\cond}{\, | \,}

\usepackage{authblk}

\author[1]{David Swanson\footnote{ORCID: 0000-0003-3174-1656}}
\author[5]{Alexander Sherry}
\author[2]{Cara Haymaker}
\author[3]{Alexandre Reuben}
\author[4]{Chad Tang}
\affil[1]{Department of Biostatistics,}
\affil[2]{Department of Translational Molecular Pathology,}
\affil[3]{Departments of Thoracic/Head and Neck Medical Oncology,}
\affil[4]{Department of Radiation Oncology, \hspace{-4cm}  \newline University of Texas MD Anderson Cancer Center;}
\affil[5]{Department of Radiation Oncology, Mayo Clinic} 

\date{December 2025}

\title{Identifying expanding TCR clonotypes with a longitudinal Bayesian mixture model and their associations with cancer patient prognosis, metastasis-directed therapy, and VJ gene enrichment}

\begin{document}

\maketitle

\abstract{Examination of T-cell receptor (TCR) clonality has become a way of understanding immunologic response to cancer and its interventions in recent years.  An aspect of these analyses is determining which receptors expand or contract statistically significantly as a function of an exogenous perturbation such as therapeutic intervention.  We characterize the commonly used Fisher's exact test approach for such analyses and propose an alternative formulation that does not necessitate pairwise, within-patient comparisons. We develop this flexible Bayesian longitudinal mixture model that accommodates variable length patient followup and handles missingness where present, not omitting data in estimation because of structural practicalities. 
Once clones are partitioned by the model into dynamic (expanding or contracting) and static categories, one can associate their counts or other characteristics with disease state, interventions, baseline biomarkers, and patient prognosis. 
We apply these developments to a cohort of prostate cancer patients who underwent randomized metastasis-directed therapy or not.  Our analyses reveal a significant increase in clonal expansions among MDT patients and their association with later progressions both independent and within strata of MDT.  Analysis of receptor motifs and VJ gene enrichment combinations using a high-dimensional penalized log-linear model we develop also suggests distinct biological characteristics of expanding clones, with and without inducement by MDT.

\noindent {\bf keywords:} T-cell receptor, VDJ genes, motif analysis, Bayesian mixture model, penalized log-linear model
}



\section{Introduction}



The adaptive immune system plays a central role in human health in part by recognizing and reacting to pathogens present in the body.  A critical component of it is T-cells, whose eponymous receptors can achieve extraordinary diversity through non-homologous recombination, with some estimates in the range of achieving $10^{18}$ different T-cell receptor (TCR) genes \cite{jokinen_predicting_2021}.  Analyses of the TCR repertoire has grown more common in recent years as technology for it has developed and their association with immunologic function has been studied and recognized as important \citep{valpione_immune_2020,kidman_characteristics_2020,bensouda_koraichi_inferring_2023,chen_human_2024}.  

While the potential and realized diversity of TCRs is extreme, only a subset of them incur expanding or contracting behavior during immune response.  Because statistical noise is a central problem in any setting with data, deciding what constitutes an expanding, contracting, or more generally, dynamic (when there are more than 2 longitudinal cross-sections), clone is an important analysis question generally and with respect to prognosis in cancer \citep{chen_longitudinal_2022,guo_characteristics_2020}.  Our first contribution in this paper is proposal of a longitudinal Bayesian mixture model consisting of two components, one which fits expanding or contracting longitudinal clonal frequencies, and one which fits static.  After fitting the model by sampling the large parameter space with Hamiltonian Monte Carlo (HMC), one is able to make probabilistic statements on whether a particular clone exhibits expanding, contracting, or static behavior in baseline-followup analyses, or dynamic versus static behavior generally \citep{betancourt_hamiltonian_2015,girolami_riemann_2011}.  In keeping with tradition in longitudinal statistics, in proposing the model we seek greater generality than the Fisher's exact and Beta-binomial approaches commonly used in TCR analyses \citep{fisher_statistical_1928,fisher_interpretation_1922,rytlewski_model_2019}, which are primarily suited to pairwise and fully-observed settings, in contrast to our approach which accommodates missing at random (MAR) mechanisms and arbitrary followup \cite{laird_random-effects_1982,little_statistical_2019,little_missing_2024}.  We also note that probabilistic statements on membership in one or the other dynamic versus static categories afforded and yielded by a mixture model accommodate flexible thresholding in associating dynamic clones with patient outcomes--downstream analysis may show that the number of high-probability expanding clones are associated with patient outcomes, while the analogous number using a less stringent threshold are not.  

One reason for receptor diversity is their constitution of V, D, and J genes. Some analyses of V, D, and J gene family interaction or enrichment have been performed in the literature, that is examination of combinations of gene families that occur together in higher or lower proportion than expected given marginal totals, but most perspectives on the topic take a visual or descriptive approach \citep{ma_analyzing_2016,shugay_vdjdb_2018,wang2025biased,chen2025characteristics}.  
An important gap to address in VDJ enrichment analyses is therefore how to calculate statistical significance of coincidence patterns of the genes that present across samples, within samples exposed to certain interventions, or, to incorporate the modeling framework outlined above, within groups of TCRs showing distinct expansion and contraction behavior.  

An established class of models within categorical data analysis and ones well-positioned to assess statistically significant co-enrichment of VJ gene families are log-linear models \citep{agresti2013categorical}.  They are a generalization of several model classes like logistic regression and can encode flexible interactions within contingency tables of arbitrary shape and number of categories $K_1\times K_2 \times \cdots \times K_N$.
Because one generally may approach log-linear modeling with hypotheses and structure one has from expert knowledge, VJ enrichment presents the challenge of necessitating a hypothesis-free and flexible model space.  In using a fully-saturated model parameterization, we therefore introduce a L1 penalization parameter which is learned from the data to achieve a sparse model fit by shrinking to zero a large subset of the parameters in the excessively large and flexible model space \citep{tibshirani1996regression}.  This framework constitutes the second methodological contribution of the manuscript.  

We begin the exposition in Sections \ref{sec:error} and \ref{sec:alternative} by analyzing and contrasting error models of the Fisher's exact test approach and our model in distinguishing static from expanding/contracting clones.  In Section \ref{sec:model} we specify the Bayesian mixture model proposed as an alternative.  We conclude our Methods in Section \ref{sec:penalized} with a description of the log-linear L1 penalized regression model framework for identifying joint VJ gene and patient characteristic enrichment patterns. 
In Section \ref{sec:results}, we perform an extensive study of a prostate cancer cohort to whom metastasis-directed therapy (MDT) and its lack (no MDT) had been randomized, with each arm consisting of a mixture of patients who received additional intermittent or continuous androgen deprivation therapy (ADT).  Using the tools developed in Methods, we fit our model to the longitudinal TCR frequencies to identify counts of expanding, contracting, or, generally, dynamic clonotypes.  We associate these groups and counts thereof with patient characteristics including MDT, receptor motifs, baseline biomarkers, and progression-free survival (PFS). Additional entropy and distributional analyses demonstrated trajectory differences between expanding and contracting clones.  Finally, application of the penalized log-linear model framework found several statistically significant enrichment of VJ gene families and some interaction and clonotype expansion status.  
We conclude in Section \ref{sec:discussion} with a discussion of the clinical and biological insights offered by our analysis, possible therapeutic targets, and ways in which one could improve the Bayesian mixture and penalized log-linear models.







\section{Methods}
\label{sec:methods}

\subsection{Error model for Fisher's exact test to assess longitudinal clonal changes}
\label{sec:error}

We begin by considering the error model of the Fisher's exact test which will prove useful when considering the expanding and contracting clonal trajectories the test is powered to detect.  Consider the $i^{th}$ clone template count for person $j$ at time $k$, $C_{ijk}$, and the total number of template counts for person $j$ at time $k$, $O_{jk} = \sum_{i=1}^{U_{jk}} C_{ijk}$, where $U_{jk}$ is the number of unique clones for person $j$ at time $k$. Since there is significant variability in the total number of template reads and it is of primary biological interest how proportions of clones change over time, we model the proportion, or frequency, $C_{ijk}/O_{jk}$ over time and patient.  


In the Fisher's exact approach, one takes pairs of these measurements along longitudinal followup and performs hypothesis tests for changes in them  \cite{dewitt_dynamics_2015,fisher_interpretation_1922} and likewise for a variation thereof with a weakly informative Beta prior \cite{rytlewski_model_2019}.  These are reasonable approaches that have been used with success, and one can attempt to understand their statistical characteristics.  Consider such a test when we have observed $C_{ij1}$ and $C_{ij2}$, that is the template count of clonotype $i$ of person $j$ at times 1 and 2 respectively and without loss of generality, and correspondingly $O_{j1}$ and $O_{j2}$, the total reads at those times.  

The null hypothesis that the proportion of the $C_{ijk}$ to $O_{jk}$ is constant over $k=1,2$ is tested with the hypergeometric distribution.  So envisioning these 4 cell counts $C_{ijk}$ and $O_{jk} - C_{ijk}$ for $k=1,2$ as a $2\times 2$ table (Table \ref{tab:conting}) and conditioning on all marginal totals, one can calculate the Fisher's exact test p-value.  


\begin{table}[ht]
\centering
\begin{tabular}{c|c|c|}
&Time 1 &Time 2\\
\hline 
clone $ij$ & $C_{ij1}$ & $C_{ij2}$ \\ 
\hline 
not clone $ij$ &$O_{j1}-C_{ij1}$ & $O_{j2}-C_{ij2}$ \\ 
\hline 
\end{tabular}
\caption{$2 \times 2$ contingency table of TCR template reads and total number of reads for person $j$ at times $k=1,2$}
\label{tab:conting}
\end{table}

It is known that the Fisher's exact test is approximated increasingly well 
by the usual score test of equal proportions as cell counts increase and convergence to some proportion $C_{ijk}/O_{jk}\rightarrow p_k$ for $k=1,2$, different in that one is only conditioning on 2 cell count marginal distributions along the vertical axis as opposed to additionally the horizontal axis \citep{casella_statistical_2024}.  For convenience of that form and exposition, we therefore consider the test statistic ${\hat p}_1=C_{ij1}/O_{j1}$ and ${\hat p}_2=C_{ij2}/O_{j2}$ and ${\hat p}=(C_{ij1}+C_{ij2})/(O_{j1} + O_{j2})$

$$\frac{{\hat p}_1 - {\hat p}_2}{\sqrt{(1/O_{j1} + 1/O_{j2})\cdot{\hat p}(1-{\hat p})}} \sim N(0,1)\mbox{ under }H_0$$
which follows a standard normal distribution under the null hypothesis \cite{rosner_fundamentals_2006}. One notes the variance estimator in the denominator, which is powered as a function of the total count of template reads at person-time $jk$, and importantly whose form is $p(1-p)$.  For small $p$, which low frequency clonotypes are, $1-p \approx 1$ so that the unit variance is approximately $p\approx p(1-p)$. However, for higher frequency clonotypes, one cannot ignore the $1-p$ term, and so the variance is $O(p^2)$.  Since $p<1$, this implies a smaller variance and therefore standard deviation.
So higher frequency clones require less proportional change to be viewed as a significant expansion or contraction because of their smaller denominator in the test statistic, other things equal.  This phenomenon occurs over and above the decrease in error coverage and therefore increase in power for some fixed longitudinal change as the mean scales linearly with $\lambda$ while the standard deviation increases at square root rate.  
This may be an appropriate approach in some cases because incorporation of the biological prior of greater interest in higher frequency clones may be warranted. However, it may not duly acknowledge biological variability that one may want to specify as more nearly proportional to the clonal mean frequency so that one does not label trivial biological change as statistically significant dynamic behavior--that is, one does not want to detect a change from, say, $11$ to $11.5$ percent frequency as a significant change even if a change from $0.5\%$ to $1.0\%$ should be.

\subsection{Alternative error model}
\label{sec:alternative}

It may seem preferable to keep a consistent view of the error term and therefore what constitutes significant biological variability by using the model $C_{ijk} \sim \mbox{Pois}(\lambda \cdot O_{jk})$, $O_{jk}$ treated as an offset, for some $\lambda$ whose distribution has the important property $E[C_{ijk}] = Var[C_{ijk}]= \lambda \cdot O_{jk}$.  Note that because error coverage of a random variable scales with the standard deviation rather than variance, this formulation still incorporates a biological prior of greater interest in higher frequency clones,
while maintaining a more consistent perspective on the error term.  This means that one is better powered for a proportional change in some clone at higher as compared to lower frequencies.  This is because for some change proportional to the mean, call it $\delta \cdot E[C_{ijk}]$ for some $0<\delta < 1$, the inverse of the coefficient of variation, a value
associated with its powering, is $$\frac{\delta\cdot E[C_{ijk}]}{\sqrt{Var[C_{ijk}]}}=\delta \frac{\lambda \cdot O_{jk}}{\sqrt{\lambda \cdot O_{jk}}}=\delta \sqrt{\lambda \cdot O_{jk}}$$
As the expression increases in $\lambda$, which is interpreted as the expected frequency, and keeping $\delta$ constant, one is better powered to detect small changes as the mean frequency increases.

\subsection{Proposed model hierarchy}
\label{sec:model}

In the Fisher's exact test setting, the null hypothesis amounts to a common success probability $p_k$ across time $k$ so that using the notation above, we have $H_0: p_1=p_2$.  Analogous for our proposal, our model attempts to distinguish between a Poisson intensity parameter $\lambda$ that is common across time or varies. We therefore use two distinct parameterizations, one that is static, $\lambda_{ij}$, and one that changes or is dynamic across time, $\lambda_{ijk}$.  That is, we have
\begin{align}
\mbox{Dynamic: } C_{ijk} \cond O_{jk} \sim \mbox{Pois}(\lambda_{ijk} O_{jk}) \vspace{2cm}\,\,\,\,\,\,\mbox{Static: }C_{ijk} \cond O_{jk} \sim \mbox{Pois}(\lambda_{ij} O_{jk})
\end{align}
Notably, while a simple Fisher's exact test or test of proportions would be comparing 2 cross sections in time, here we index generically by some $k=1,\dots,T_j$, where $T_j$ is the number of followup times of observations $j$, some of which may be missing for $C_{ijk}$, and therefore is a more general formulation.  Because $\lambda$ is a single degree of freedom so that marginals are not sufficiently flexible to fully describe observed data and also motivated by Bayesian conjugacy, we assert the hierarchical formulation:
%
\begin{align}
\lambda_{ijk} \cond \alpha_j, \beta_j \sim \mbox{Gamma}(\alpha_j, \beta_j) \;\;\;\;\;\; \lambda_{ij} \cond \alpha_j, \beta_j \sim \mbox{Gamma}(\alpha_j, \beta_j)
\end{align}
noting that $\alpha_j$ and $\beta_j$ are held in common across the two components and that indexing them by $j$ addresses possible within-person correlation induced by variation in the number of unique clones across subjects.


Leveraging conjugacy relationships yields the following convenient probability mass functions upon integration out of $\lambda_{ij}$ or $\lambda_{ijk}$ depending on the component:
\begin{align}
p_S(C_{ij\cdot } |O_{j\cdot},  \alpha_j, \beta_j)=&\int p_S(C_{ij\cdot },\lambda_{ij} | O_{jk},  \alpha_j, \beta_j) \, d\lambda_{ij} = \int l_S(C_{i\cdot} | \lambda_{ij} \,O_{jk})\, p(\lambda_{ij} | \alpha_j, \beta_j) \, d\lambda_{ij}  \nonumber \\
=&\frac{ \Gamma(\sum_k C_{ijk} + \alpha_j) }{ \Gamma(\alpha_j) \prod_{k=1}^{T_j} C_{ijk}!}  \biggl(\frac{ \beta_j}{\beta_j + \sum_k O_{jk}}\biggr)^{\alpha_j}  \prod_{k=1}^{T_j} \biggl( \frac{O_{jk}}{\beta_j + \sum_k O_{jk}}\biggr)^{C_{ijk} }
\end{align}

\begin{align}
p_D(C_{ij\cdot } |O_{j\cdot},  \alpha_j, \beta_j)=&\int p_D(C_{ij\cdot },\lambda_{ijk} | O_{jk},  \alpha_j, \beta_j)\, d\lambda_{ijk} =\int l_D(C_{i\cdot} | \lambda_{ijk} \,O_{jk})\, p(\lambda_{ijk} | \alpha_j, \beta_j) \, d\lambda_{ijk} \nonumber \\
=&\prod_{k=1}^{T_j} \frac{ \Gamma( C_{ijk} + \alpha_j) }{ \Gamma(\alpha_j)  C_{ijk}!}  \biggl(\frac{ \beta_j}{\beta_j + O_{jk}}\biggr)^{\alpha_j}   \biggl( \frac{O_{jk}}{\beta_j + O_{jk}}\biggr)^{ C_{ijk}}
\end{align}
The mixture of dynamic and static components is then
$$p(C_{ij\cdot } |O_{j\cdot},  \alpha_j, \beta_j) = p_S(C_{ij\cdot } |O_{j\cdot},  \alpha_j, \beta_j)+ \pi \cdot \,p_D(C_{ij\cdot } |O_{j\cdot},  \alpha_j, \beta_j) $$
where the marginal or population mixing proportion for dynamic clones is $\pi$.  We show the derivation in greater detail in Section \ref{sec:derive}.  We remark that integration of $\lambda_{ij}$ and $\lambda_{ijk}$ out of their respective expressions yields a negative multinomial or product of negative binomials probability mass functions as the posterior predictive distributions for the static and dynamic components, respectively, modified according to the offset term.  Indexing $\alpha_{j}$ and $\beta_{j}$ by person adds flexibility and allows for a variable number of unique clones within patient, which is desirable because inherent in modeling frequencies is the linear dependence of their sum equaling 1.

The empirically estimated $\alpha_j$ and $\beta_j$ will play a moderating role in the ``powering'' of the mixture model, that is the ability of it to distinguish clonotypes falling into respective components.  Since they are empirically estimated based on realized frequencies, they are not subject to analyst-chosen biases.  As we proceed in examining the two component probability mass functions, it is notable that for data whose estimated $\alpha_j$ and $\beta_j$ imply a small mean and variance, one is less powered to distinguish between static and dynamic behavior precisely because of the points made above--when drawing different $\lambda_{ijk} \cond \alpha_j, \beta_j \sim \mbox{Gamma}(\alpha_j,\beta_j)$ with a small mean and variance, the Poisson error overwhelms variability at the next level of hierarchy and one cannot distinguish the two components.  Indeed, one can have an arbitrarily large mean with a small variance, but the tendency and worst case scenario will be small mean and small variance because the coverage of the Poisson error gets smaller relative to the mean for an increasing mean.

To share information across the population, we introduce another layer of hierarchy:
\begin{align}
\log \alpha_j \sim N(\mu_\alpha, \sigma^2_\alpha) \,\,\,\,\,\,\,\, \log \beta_j \sim N(\mu_\beta,\sigma^2_\beta)
\end{align}
where $\mu_\alpha$, $\mu_\beta$, $\sigma^2_\alpha$, and $\sigma^2_\beta$ will be estimated empirically, and $\log$ or $\mbox{logit}$ are applied to induce symmetry about zero on the parameter. 
We place a diffuse and weak prior on $\pi$ asserting $$ \mbox{logit}\, \pi \sim N(0, \sigma^2_{\pi})$$
for large $\sigma^2_{\pi}$ and fit the model using Hamiltonian Monte Carlo (HMC), implemented in STAN, because of the hierarchical formulation and lack of closed-form integrals \citep{hoffman_no-u-turn_2014,stan_development_team_stan_2024}.  Recovery of parameters by the fitting procedure is demonstrated in Section \ref{sec:sim}.






\subsection{Penalized log-linear modeling for VJ gene enrichment testing and interaction with patient phenotype}
\label{sec:penalized}

We conclude the Methods by proposing a different model and one serving a complementary purpose to that above which partitions expanding and contracting clones from static ones.  One aspect of TCR repertoire analysis is examination of V, D, and J gene recombination, and where there exists gene family enrichment marginally or in combinations of these gene families.  To our knowledge, a formal framework for modeling and testing statistical significance of this enrichment has not been proposed for VDJ analyses.  Here, we propose using L1 penalized and fully saturated log-linear models for parameterizing the contingency tables of gene family coincidences by which these data can be described.  When using this framework, one can introduce still other characteristics of the T-cell receptor, such as whether it expanded or contracted, and thus test for associations between gene families and that characteristic.  In the model description below, we focus on just $V$ and $J$ genes in our notation and then some arbitrary TCR or patient characteristic notated with $P$, done for exposition and to connect the notation with our analysis in Section \ref{sec:results}.

Let $G_{pvj}$ be the count of the number of clonotypes with patient characteristic $p$, $V$ gene family v, $J$ gene family j.  That is, $G_{pvj}$ is the number of $C_{ij\cdot}$ composed of those gene families and found in patient $j$ with phenotype $p$.
Then define $$f(\beta_p^P,\dots,\beta_{pvj}^{PVJ})=f(\beta_p^P,\beta_v^V,\beta_j^J,\beta_{pv}^{PV},\beta_{pj}^{PJ},\beta_{vj}^{VJ},\beta_{pvj}^{PVJ}) = \beta_p^P+ \beta_v^V+ \beta_j^J  + \beta_{pv}^{PV}+ \beta_{pj}^{PJ} + \beta_{vj}^{VJ}+ \beta_{pvj}^{PVJ} $$
and now consider the model 
\begin{align}
\log \, {G}_{pvj} = \beta + f(\beta_p^P,\dots,\beta_{pvj}^{PVJ}) + \lambda_\beta \sum_{pvj} f^{|\;|}(\beta_p^P,\dots,\beta_{pvj}^{PVJ})
\end{align}
where $f^{|\;|}(\beta_p^P,\dots,\beta_{pvj}^{PVJ})= |\beta_p^P|+ |\beta_v^V|+ |\beta_j^J|  +| \beta_{pv}^{PV}|+ |\beta_{pj}^{PJ}| +| \beta_{vj}^{VJ}|+ |\beta_{pvj}^{PVJ} |$
and $\lambda_\beta$ is the penalization parameter, and this adopts the notation of log-linear models for modeling contingency tables of arbitrary size and dimension. 
So here the interpretation of, say, $\beta_{pv}^{PV}$ is the odds ratio of patient characteristic $p$ for having V gene $v$ as compared to the V family gene reference category.  This is a saturated and high-dimensional, L1 penalized log-linear model, and, because of the penalty, one achieves sparse solutions under the hypothesis that only a subset of VJ interactions, indicating enrichment, with or without patient phenotype are significant.  The fully saturated parameterization means it has $N_P\cdot N_V\cdot N_J$ parameters, the respective numbers of categories of patient characteristics, v genes, and j genes, respectively, so that every cell in the contingency table can be modeled without error prior to penalization.
We estimate $\lambda_\beta$ by cross validation.  After identifying a sparse subset of parameters on which to fit the model, we do so and identify significance of parameters in the usual way.



\section{Results}
\label{sec:results}


We fit the longitudinal Bayesian mixture model to 62662 productive T-cell receptors observed over baseline and followup among 97 prostate cancer patients randomized to MDT and no MDT, and to 86,381 productive TCRs on 104 observations for the baseline with two followups analysis.  Results shown are for the baseline-followup analysis unless otherwise noted, for example as in the dynamic clone by MDT stratum comparison, out of consideration of missing at random (MAR) assumptions necessary in the baseline with two followups analysis.
 T-cell receptor–$\beta$ complementarity-determining region (CDR) 3 regions were sequenced with an immunoSEQ assay of Adaptive Technologies, where peripheral blood samples were derived from patients at their visits.   
 We filtered T-cell receptors for non-productive sequences and those with $<$8 total read counts across all of followup so that if a clone had few reads at one cross section, but experienced significant expansion, it would still be included in analysis. The threshold was chosen to filter those clones of likely little biological relevance in addition to numerical artifacts arising in the mixture when modeling low counts.  After convergence of the Hamiltonian Monte Carlo fitting procedure \citep{betancourt_hamiltonian_2015,stan_development_team_stan_2024}, we counted the number of expanding, contracting, and static clones within each patient based on some threshold probability of dynamic component membership, which we chose
as 0.95 so that the more aggressively dynamic, expanding, or contracting clones would be labeled as such, having observed stronger associations with a more stringent definition.  



\subsection{Prognostic models}

We fit Cox proportional hazards models to the progression-free survival (PFS) endpoint in the prostate cancer cohort, with 53 events observed on the 97 subjects with the immunoSEQ assay at their baseline and followup visits \cite{cox1972regression,ludmir_addition_2024,sherry_definitive_2022}.  We log-transformed the counts of the number of expansions and contractions because of the right-skewness of the measure and to make inference robust to the influence of leveraged observations.  To improve estimation efficiency and account for disease and performance status heterogeneity, we adjusted the Cox model for patient age, ADT, and number of lesions.  

The three Cox models shown in Table \ref{tab:cox} demonstrate that clonal expansions tend to be more significant in prognostic models than dynamic clones (expansions and contractions combined) or contractions, and more expansions suggest protection against a PFS event, though significance depends on whether the model is adjusted for MDT, and one would expect to see greater significance in the model with MDT were one better-powered.  MDT is the most prognostic and protective of explanatory variables in time-to-event analysis.  We also see that expansions tend to be more significant with inclusion of contractions, suggesting that it is within strata of contractions that expansions are more protective.  The interpretation is that the ratio of these is important for explaining variation in PFS, though formal testing of this hypothesis is difficult and underpowered and did not yield a significant result.  Lastly, MDT confounds the expansions-PFS relationship, though examination of parameter estimates suggests modest mediation of MDT-PFS by the number of expanding clones.  

\begin{figure}[H]
\centering
\includegraphics[width=0.65\textwidth]{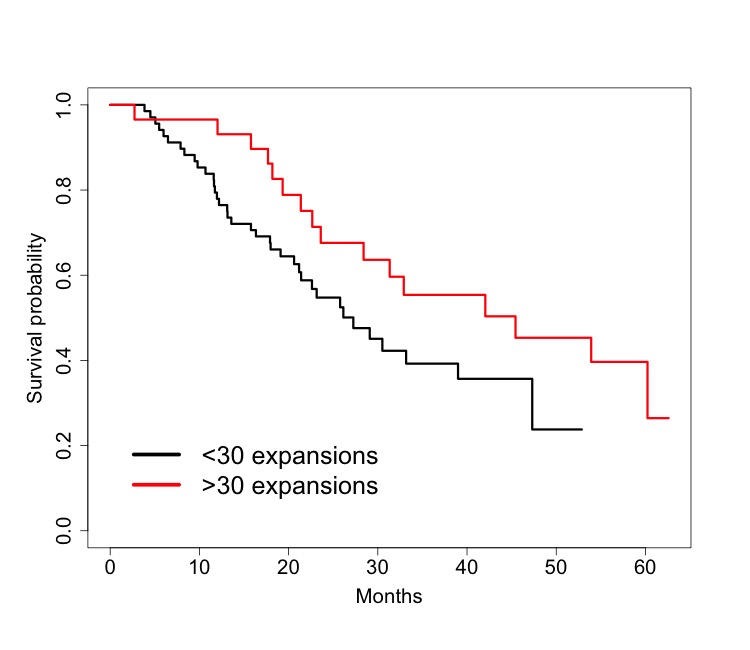} 
\caption{KM of PFS stratified by number expansions.  Log-rank test of the dichotomized strata yields a p-value of 0.08, and when treated as continuous in a Cox model 0.06. } 
\label{fig:km}
\end{figure}

To investigate marginal associations and by way of non-parametric methods, we used the Kaplan-Meier estimator to estimate PFS stratified by greater or fewer than 30 expansions, yielding a log-rank test with p=0.08 (Figure \ref{fig:km}).  The threshold was chosen to reflect a population mode of expansions fewer than the number, and was approximately the 70th percentile of clonal expansions count.  When patient expansion count was treated as continuous in the model, we observed a slightly more significant result (p=0.06).

 \begin{table}[ht]
 \centering
 \begin{tabular}{rllll}
   \hline
  &  $\widehat{HR}$ & 95\% CI & Z-statistic & Pr($>$$|$Z$|$) \\
   \hline
 {\bf Full Model} & & & &  \\
    MDT & 0.321 & (-1.8, -0.5) & -3.51 & 0.000449 \\
   logged expansions  & 0.825 & (-0.46, 0.075) & -1.41 & 0.158 \\
   logged contractions & 1.19 & (-0.13, 0.48) & 1.15 & 0.251 \\
    {} & & & &  \\
    {\bf Dynamism only} & & & &  \\
   logged expansions  & 0.743 & (-0.56, -0.033) & -2.2 & 0.0275 \\
   logged contractions & 1.11 & (-0.19, 0.4) & 0.711 & 0.477 \\
    {} & & & &  \\
 {\bf Expansions excluded} & & & &  \\
   MDT & 0.299 & (-1.8, -0.6) & -3.87 & 0.00011 \\
   logged contractions & 1.03 & (-0.19, 0.25) & 0.268 & 0.789 \\

    \hline
 \end{tabular}
 \caption{Three PFS Cox models with combinations of expansions/contractions and MDT, all adjusted for Age, ADT, and lesion group.}
 \label{tab:cox}
 \end{table}

\subsection{Enrichment of VJ gene families and clonal expansions}

We fit the penalized saturated log-linear model to V and J gene families and whether the clone was determined as expanding, contracting, or dynamic, with results shown in Table \ref{tab:vdj}.  We did not include D genes in our analysis because information on their gene family was only available for a minority of the TCR.  Model fits suggested highly significant enrichment above expected between select V and J gene families, and no enrichment by clone contraction status.  The main effect of expansion status in Table \ref{tab:vdj} is estimated as a large negative number because expanding clones are in the minority of the clone population.  The analysis was well-powered because of the thousands of TCR analyzed and so we focus on the orthogonal information offered by the effect size parameters, $\hat{\beta}$, which are helpful to understand magnitude of difference by fold change.  Most $\hat{\beta}$'s are positive, suggesting that when there is gene family interaction, joint frequencies are greater than expected under the null of no interaction rather than less, and one observes J gene family 02 with V family 11 and 24 in particular as having some of these highest levels.  We did observe significant interaction between being an expanding clone and V gene family 06, in this case with coincidence below what would be expected by their marginal frequencies, though a biological rationale for the phenomenon was unclear.  


 \begin{table}[H]
 \centering
 \begin{tabular}{rllll}
   \hline
   & $\hat{\beta}$ & 95\% CI & Z-statistic & Pr($>$$|$Z$|$) \\
   \hline
 {\bf  Expansion Main Effect }&&&& \\
 Expansion & -2.95 & (-3, -2.9) & -144 & 0 \\
 { }&&&& \\
 {\bf  Gene family interactions }&&&& \\ 
   TCRBJ02 * TCRBV02 & -0.0321 & (-0.13, 0.066) & -0.643 & 0.52 \\
   TCRBJ02 * TCRBV03 & 0.166 & (0.055, 0.28) & 2.94 & 0.00328 \\
   TCRBJ02 * TCRBV04 & 0.203 & (0.11, 0.3) & 4.26 & 2.02e-05 \\
   TCRBJ02 * TCRBV05 & 0.202 & (0.12, 0.28) & 4.88 & 1.05e-06 \\
   TCRBJ02 * TCRBV06 & 0.114 & (0.035, 0.19) & 2.82 & 0.00483 \\
   TCRBJ02 * TCRBV07 & 0.263 & (0.18, 0.34) & 6.57 & 5.13e-11 \\
   TCRBJ02 * TCRBV09 & 0.327 & (0.22, 0.43) & 6.04 & 1.59e-09 \\
   TCRBJ02 * TCRBV10 & 0.163 & (0.047, 0.28) & 2.76 & 0.00585 \\
   TCRBJ02 * TCRBV11 & 0.547 & (0.44, 0.65) & 10 & 1.34e-23 \\
   TCRBJ02 * TCRBV12 & -0.0939 & (-0.19, 0.0043) & -1.87 & 0.0609 \\
   TCRBJ02 * TCRBV14 & 0.191 & (0.036, 0.35) & 2.41 & 0.0161 \\
   TCRBJ02 * TCRBV16 & -0.557 & (-1, -0.13) & -2.54 & 0.0112 \\
   TCRBJ02 * TCRBV19 & -0.227 & (-0.32, -0.13) & -4.71 & 2.45e-06 \\
   TCRBJ02 * TCRBV20 & 0.167 & (0.08, 0.25) & 3.74 & 0.000182 \\
   TCRBJ02 * TCRBV24 & 1.57 & (1.4, 1.7) & 20.9 & 1.19e-96 \\
   TCRBJ02 * TCRBV27 & 0.242 & (0.14, 0.35) & 4.51 & 6.41e-06 \\
   TCRBJ02 * TCRBV28 & -0.0553 & (-0.16, 0.049) & -1.04 & 0.299 \\
   TCRBJ02 * TCRBV29 & 0.129 & (0.00018, 0.26) & 1.96 & 0.0497 \\
   TCRBJ02 * TCRBV30 & -0.395 & (-0.52, -0.27) & -6.07 & 1.26e-09 \\
{ }&&&& \\
{\bf Gene family \phantom{asdsdfasdfasd} }&&&& \\
  {\bf  expansion interaction }&&&& \\ 
   TCRBV06 * Expansion & -0.168 & (-0.29, -0.048) & -2.71 & 0.00678 \\
    \hline
 \end{tabular}
 \caption{VJ gene with expansion interaction identified by log-linear model penalized regression.  Intercept and gene family main effects not shown.}
\label{tab:vdj}
 \end{table}


\subsection{Clonal expansion and contraction distributions}

\begin{figure}[H]
\centering
\begin{subfigure}[h]{0.4\linewidth}
\includegraphics[width=0.99\textwidth]{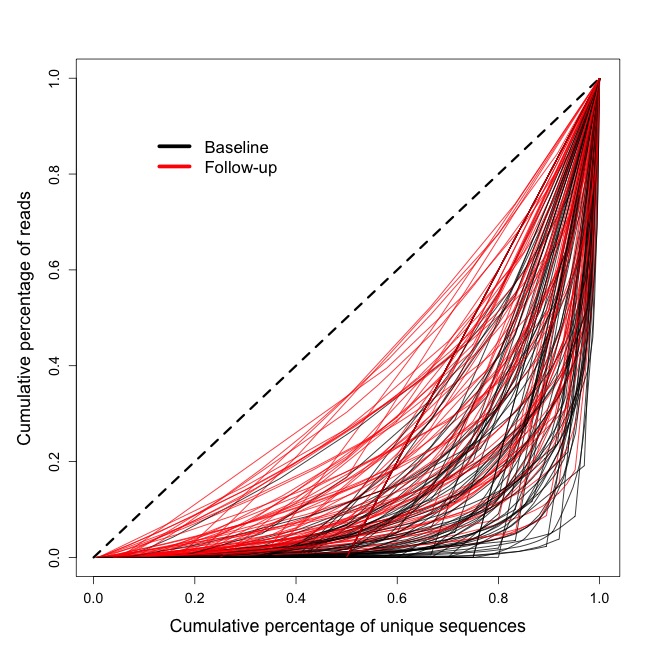} 
\caption{Expansion clones Lorenz curves.}
\end{subfigure}
\begin{subfigure}[h]{0.4\linewidth}
\includegraphics[width=0.99\textwidth]{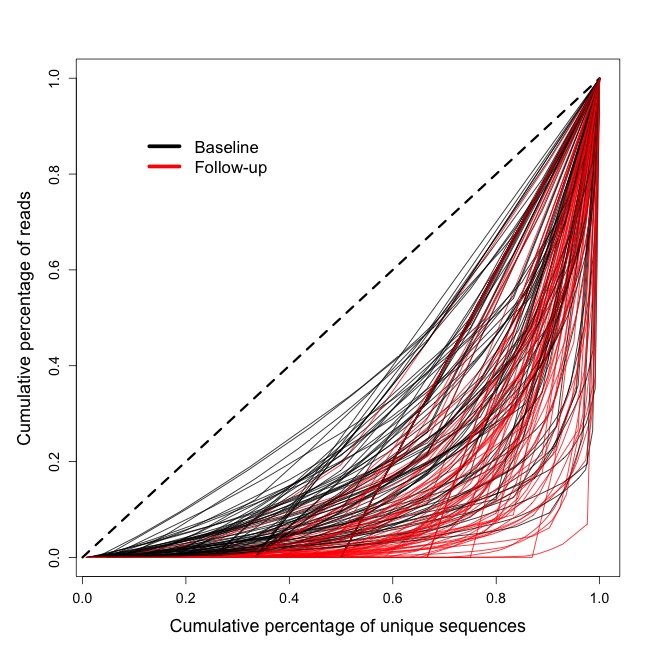} 
\caption{Contraction clones Lorenz curves.}
\end{subfigure}
\begin{subfigure}[h]{0.4\linewidth}
\includegraphics[width=0.99\textwidth]{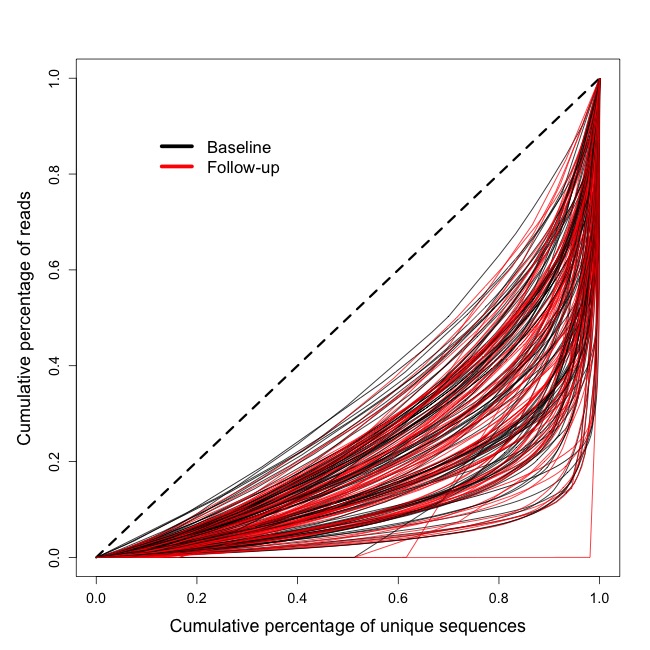} 
\caption{Static clones Lorenz curves.}
\end{subfigure}
\caption{Lorenz curves for subjects stratified by baseline/followup cross-sections for expanding, contracting, and static clones}
\label{fig:lorenz}
\end{figure}

When considering longitudinal movements of TCR clonotypes, one can examine whether expanding and contracting clones do so by some scaling factor or are translated by an absolute amount.  Scaling might suggest a strong immune response maintains existing relative proportions for some subset of clones, whereas translations by some absolute amount would tend in the limit to push those clones toward the same frequency.  There are different measures related to frequency distribution entropy including Gini coefficients, and in Figure \ref{fig:lorenz} we use and plot the Lorenz curves from which they are derived, doing so for expanding, contracting, and static clones, stratified by patient and longitudinal followup time.  The implication of the marked stratification by baseline-followup among the curves, with movement toward more evenly-distributed frequencies for the expansions, is that clonal movements on average tend to be translations as opposed to scalings--a clone starting at a higher or lower frequency will change by X amount rather than X percent.  






\subsection{Baseline biomarkers and clonal dynamism}

\begin{figure}[H]
\begin{subfigure}[h]{0.45\linewidth}
\includegraphics[width=1.0\textwidth]{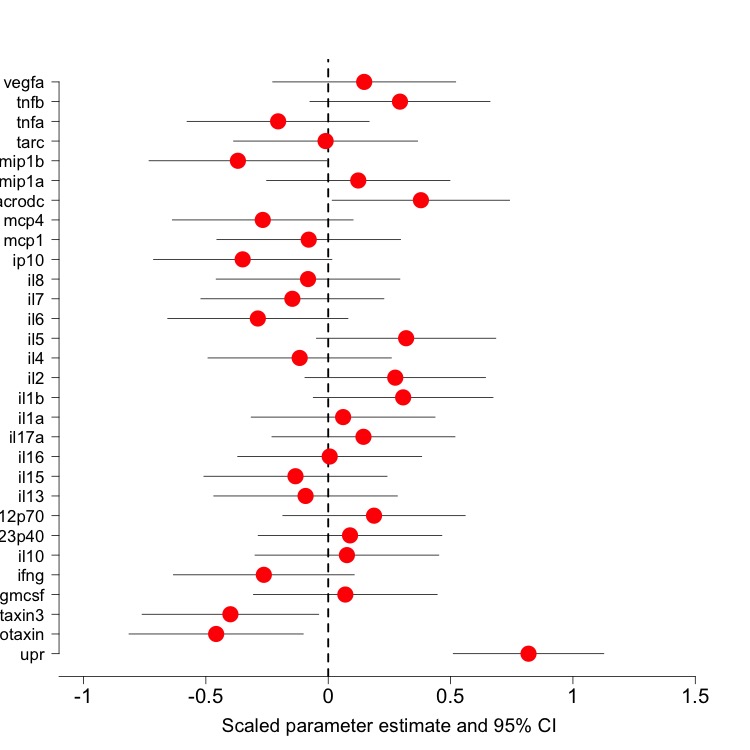} 
\caption{Marginal associations between baseline biomarkers and expansions among MDT-treated observations.}
\end{subfigure}
\hspace{1cm}
\begin{subfigure}[h]{0.45\linewidth}
\includegraphics[width=1.0\textwidth]{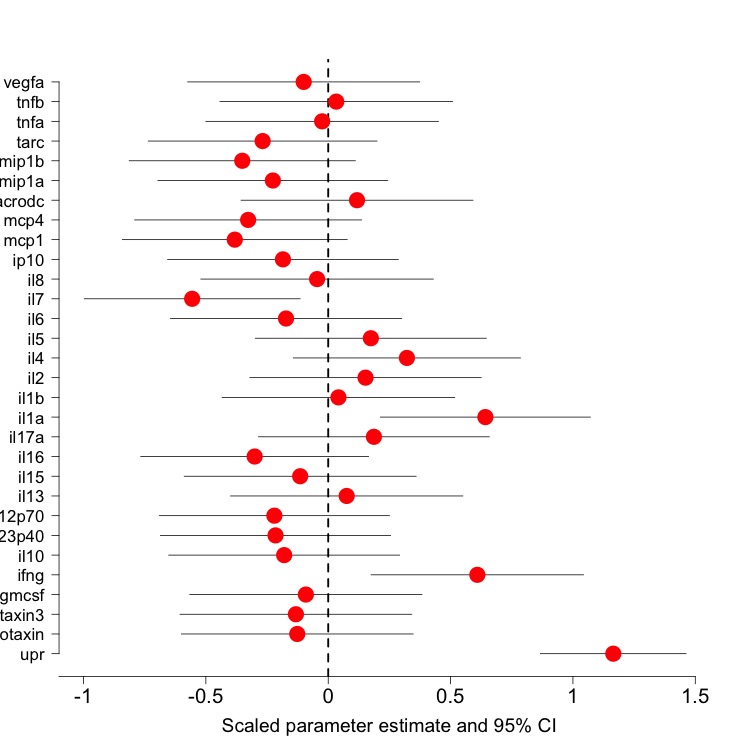} 
\caption{Marginal associations between baseline biomarkers and expansions among non-MDT observations.}
\end{subfigure}
\caption{Marginal associations between baseline biomarkers and clonal expansions, stratified by MDT.}
\label{fig:norm}
\end{figure}
\begin{figure}[H]
\begin{subfigure}[h]{0.45\linewidth}
\includegraphics[width=1.0\textwidth]{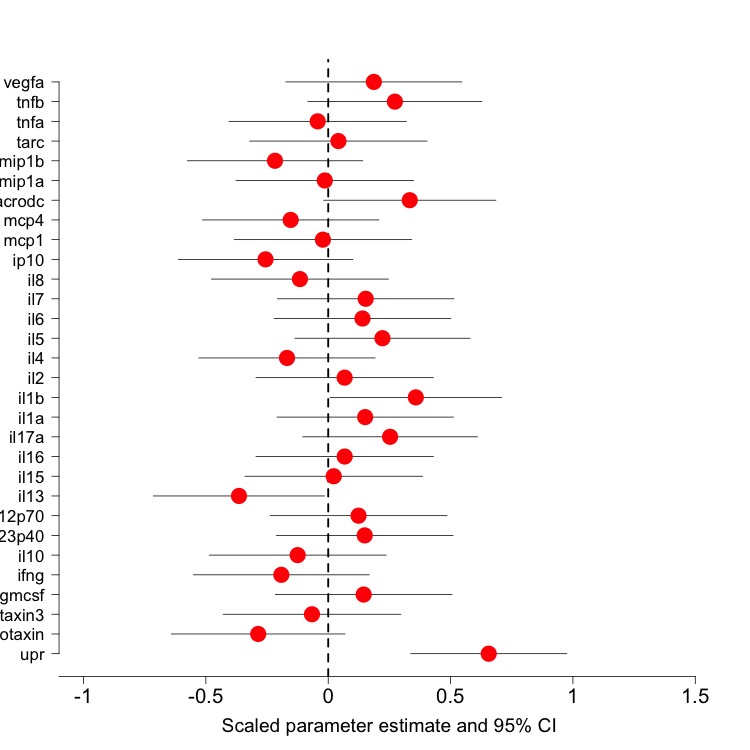} 
\caption{Marginal associations between baseline biomarkers and contractions among MDT-treated observations.}
\end{subfigure}
\hspace{1cm}
\begin{subfigure}[h]{0.45\linewidth}
\includegraphics[width=1.0\textwidth]{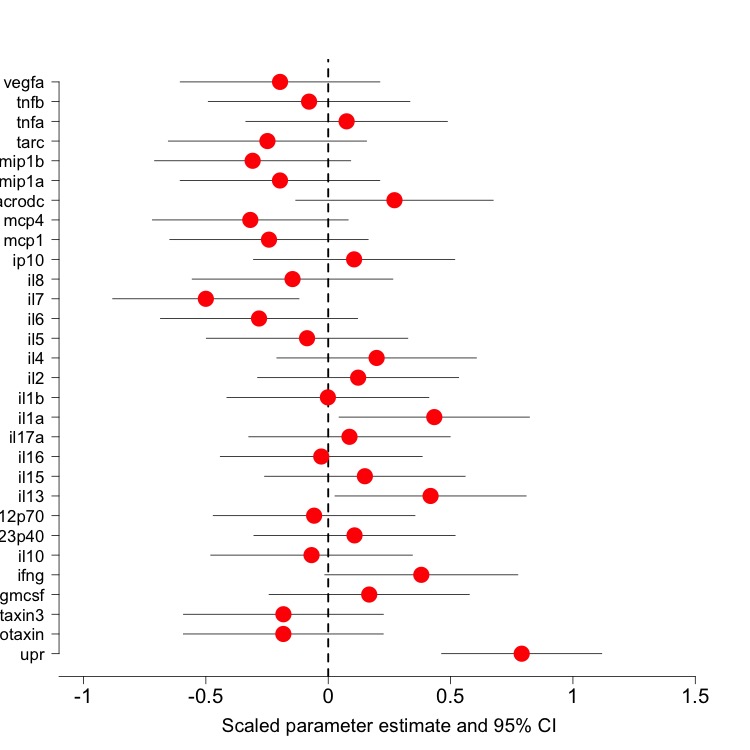} 
\caption{Marginal associations between baseline biomarkers and contractions among non-MDT observations.}
\end{subfigure}
\caption{Marginal associations between baseline biomarkers and clonal contractions, stratified by MDT.}
\label{fig:norm2}
\end{figure}

We fit linear regression models associating counts of expansions and contractions with baseline biomarkers, within strata of MDT and non-MDT patients.  Normalized marginal association parameter estimates are shown in Figures \ref{fig:norm} and \ref{fig:norm2}.  The primary, clear association between clonal dynamism and biomarkers is UPR, which doesn't vary significantly by expansion/contraction nor MDT/non-MDT status. Other, weaker associations were found (eg, Il7), but they did not pass multiple testing correction.  There was also some modest variation in the normalized effect sizes by expansion and MDT strata.

\begin{figure}[H]
\begin{subfigure}{1\linewidth}
\centering
\includegraphics[width=0.75\textwidth]{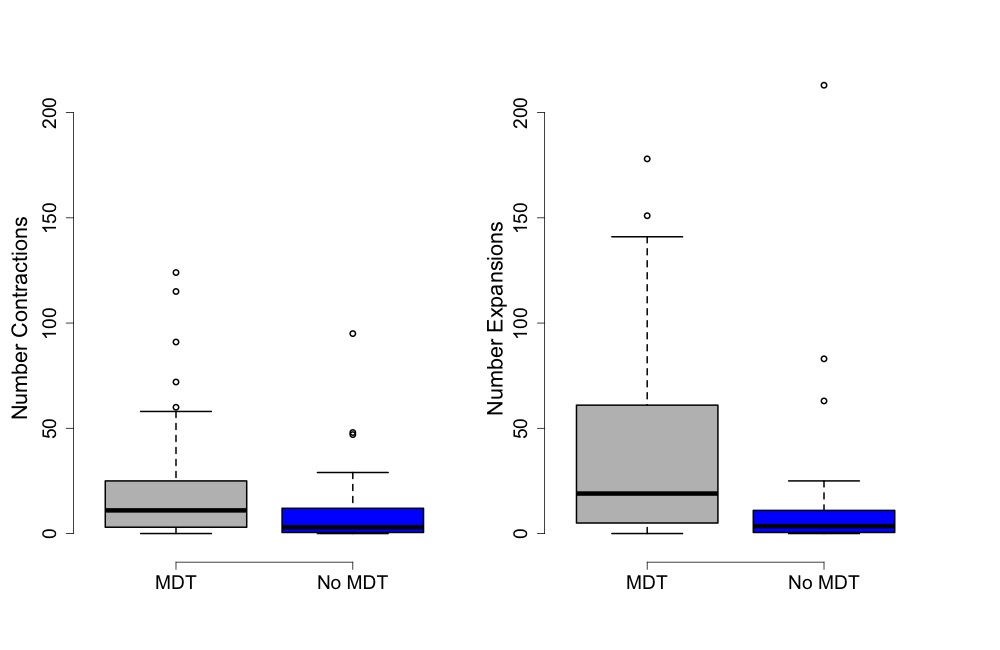} 
\caption{Contractions (LHS) and expansions (RHS) boxplots stratified by MDT for 2 followup analysis.}
\end{subfigure}
\begin{subfigure}{1\linewidth}
\centering
\includegraphics[width=0.5\textwidth]{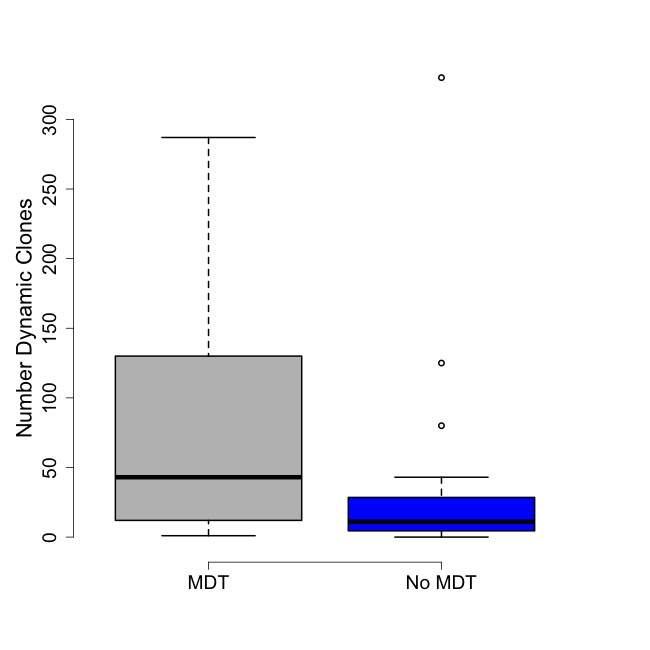} 
\caption{Dynamic clone boxplots of 3 followup analysis, stratified by MDT.}
\end{subfigure}
\caption{Expanding, contracting, and dynamic clones for 2 and 3 followup analysis, respectively.  All p-values for MDT strata $<10^{-8}$ in a generalized linear model.}
\label{fig:exp}
\end{figure}

We explored interactions of baseline biomarkers with glinternet, a model and tool for identifying these terms in penalized high-dimensional regression settings which maintains hierarchical model ordering with main effects \citep{lim2015learning}.  Results are given in Table \ref{tab:glin} and show two modest interactions between gmcsf\,*\,mip1b and il15\,*\,vegfa with respect to the number of expansions in subjects \citep{tibshirani1996regression}.  Lastly, in Figure \ref{fig:exp}, we see the MDT stratum exhibit a significantly greater number of both contractions and expansions in the baseline-followup analysis and dynamic clones generally in the baseline with two followup analysis.  We also found positive and negative associations between the presence of HLA alleles A*03 and C*07, respectively, and clonal expansions for the subset on whom HLA typing was available, with results presented in Section \ref{sec:hla}.



 \begin{table}[H]
 \centering
 \begin{tabular}{rllll}
   \hline
  & $\hat{\beta}$ & 95\% CI & t-statistic & Pr($>$$|$t$|$) \\
   \hline
   {\bf  Main effects \phantom{asdfaasdf} }&&&& \\ 
   upr & 2.93e-05 & (2.2e-05, 3.6e-05) & 8.28 & 1.42e-12 \\
   gmcsf & 7.8 & (-0.62, 16) & 1.84 & 0.0689 \\
   il10 & 0.869 & (0.092, 1.6) & 2.22 & 0.0289 \\
   il1a & 0.0236 & (-0.017, 0.064) & 1.17 & 0.247 \\
   il15 & -0.516 & (-0.94, -0.09) & -2.41 & 0.0181 \\
   vegfa & -0.0351 & (-0.067, -0.0033) & -2.19 & 0.0311 \\
   mip1b & 0.00516 & (-0.0054, 0.016) & 0.971 & 0.334 \\
    \phantom{asdfaasdf} &&&& \\ 
      {\bf  Interaction \phantom{asdafasdf} }&&&& \\ 
   gmcsf\,*\,mip1b & -0.1 & (-0.2, -0.0053) & -2.1 & 0.0388 \\
   il15\,*\,vegfa & 0.0119 & (0.002, 0.022) & 2.39 & 0.0192 \\
    \hline
 \end{tabular}
 \caption{Regressing clonal expansions on baseline biomarkers with interaction-oriented penalized regression enforcing hierarchically-ordered models}
 \label{tab:glin}
 \end{table}

\subsection{Motif and nucleotide analysis}

\begin{figure}[H]
\begin{subfigure}[h]{0.45\linewidth}
\includegraphics[width=1.0\textwidth]{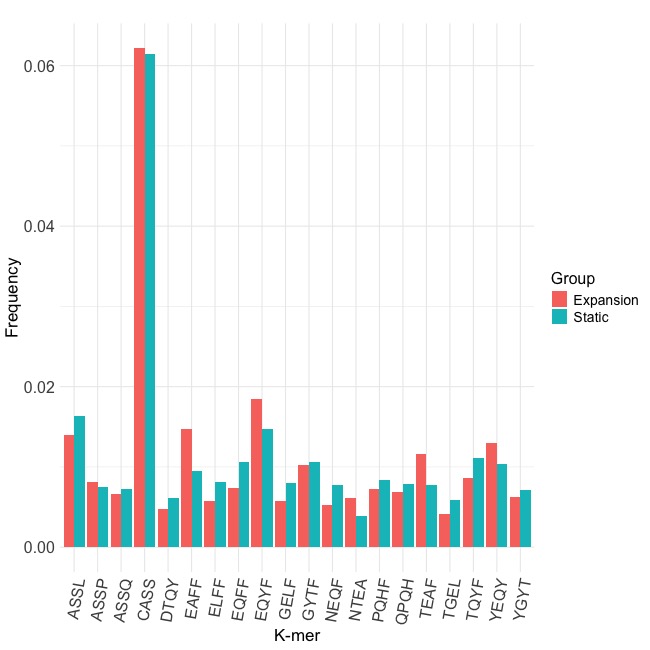} 
\caption{Most common motifs among dynamic clones, stratified by being an expansion/static clone.}
\end{subfigure}
\hspace{1cm}
\begin{subfigure}[h]{0.45\linewidth}
\includegraphics[width=1.0\textwidth]{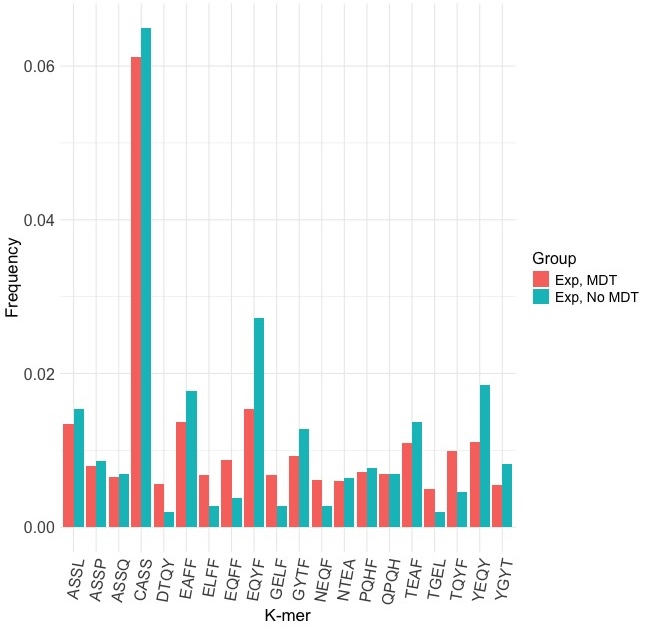} 
\caption{Most common motifs among expanding clones, stratified by being an expansion within an MDT versus non-MDT patient.}
\end{subfigure}
\caption{Common motifs stratified by expansion status and MDT.}
\label{fig:motif}
\end{figure}

Motif analysis yielded visual indication of similar relative proportions of common motifs among expanding and static clones (Figure \ref{fig:motif}a), suggesting there was not enrichment of one or another motif among expanding clones. In contrast, visual inspection of relative proportions of motifs among expanding clones stratified by MDT/non-MDT treatment status suggested differences, with greater uniformity among motifs of MDT-induced expansions.  The result may support the idea that expanding clones induced by MDT have some different motif characteristics than those present without MDT.  Formal tests of differences in distributions were not significant, but the large number of motif categories underpowers the analysis.  Despite being well-powered, we found no difference in CDR3 length between expanding, contracting, and static clones with respect to clonotype clustering.  
Likewise, there were only modest differences in average nucleotide edit distance
between expanding and static clones, matching within person for the number of clones being compared, by a t-test and Wilcoxon rank-sum (p=0.25), with those expanding having a slightly greater average edit distance. 






\section{Discussion}
\label{sec:discussion}

In this work, we explored how the TCR repertoire and its longitudinal changes are influenced by MDT and in turn have bearing on patient response as measured by progression-free survival independent of and in combination with MDT. We have also tried to characterize the nature of longitudinal clonal trajectories, be they translations or scalings, and explored how determination as a clonal expansion or contraction may be related to enrichment of particular gene families, baseline biomarkers, and receptor motifs.

An important and confirmatory insight of our analyses is that clonal expansions may have some association with patient prognosis both marginally and, weaker, within strata of MDT, contractions, and adjusted for measures of comorbidity.  The suggestion is that patients with a stronger or broader immune response with respect to the number of expanding clonotypes may be associated with later disease progression.   Because one of the clearer signals from our study is that randomized MDT significantly increases the number of expanding clones, interpretation of the modest association with PFS should be viewed with caution as it is confounded by MDT.  While our study constitutes a relatively large prospective cohort of TCR sequenced individuals, we are still underpowered on 53 PFS events from 97 patients and so analyses on future, larger cohorts are warranted.  

Much important methodological work has focused on VDJ genes, but our proposal takes an important step toward systematizing assessment of enrichment between VJ gene families in the T-cell receptor.  Because of the generality of our framework, one can further analyze interaction with other characteristics of the clonotype such as whether it expanded, contracted, or was static. Our analyses revealed interesting enrichment between several V and J gene families in addition to interaction between expanding clones and one particular V gene family. This latter result in combination with visual suggestion from Figure \ref{fig:motif}b implies there exist subtle differences between the clonotypes expanding in response to MDT versus in the absence of it.  



\section{Supplementary material}
\label{sec:supplementary}

The supplementary material includes simulation results in Section \ref{sec:sim} confirming parameter estimates align with those under which data are generated, derivation of mixing components for the longitudinal model in Section \ref{sec:derive}, and HLA typing analysis in Section \ref{sec:hla}.

\bibliographystyle{plainnat} 
\bibliography{bioinf_tcrMay23}

\pagebreak

\clearpage

\begin{center}
\textbf{\large 
Supplementary Material: Identifying expanding TCR clonotypes with a longitudinal Bayesian mixture model and their associations with cancer patient prognosis, metastasis-directed therapy, and VJ gene enrichment}

\textbf{David Swanson, Alexander Sherry, Cara Haymaker, Alexandre Reuben, Chad Tang}
\end{center}
\setcounter{equation}{1}
\setcounter{figure}{0}
\setcounter{section}{0}
\setcounter{table}{0}
\setcounter{page}{1}
\makeatletter
\renewcommand{\thetable}{A\arabic{table}}
\renewcommand{\thefigure}{A\arabic{figure}}
\renewcommand{\thesection}{A\arabic{section}}
\renewcommand{\bibnumfmt}[1]{[A#1]}
\renewcommand{\citenumfont}[1]{#1}
\newcommand{\ma}[1]{\mathbf{#1}}

\section{Simulation}
\label{sec:sim}

We performed several simulations to examine whether the fitting procedure recovers the parameters under which samples were generated.  We considered the 100 samples of logged values of $\alpha_i$ and $\beta_i$ as distributed Gaussian with mean $-0.5$ or $-0.3$ and $6.0$ or $4.4$, respectively.  The standard deviations under these parameterizations were 0.5 or 0.3, respectively.  Under subsequent exponentiation, these parameters yield approximately the same mean for the Gamma prior, but with the pairing of -0.3 and 4.4 having an on average higher variance from which the Poisson mean parameter is sampled.  Intuitively this should yield more information longitudinally with respect to the clone's dynamic or static component membership. We generated 60,000 total clones on 2 or 3 followup times depending on the simulation, 20 percent of which were generated under the dynamic model and 80 percent under the static.  For the 3 followup simulation, approximately 34\% of samples had some missingness at the 2nd or 3rd followup time.  Offset terms were generated under a Gamma distribution with shape parameter of 5000/20 and rate parameter 1/20.

The model was fit with combinations of 2 or 3 followups, in the latter case with missingness, and the $\alpha_j$ and $\beta_j$ parameterizations of the Gamma distribution with smaller or bigger variance.  We used Hamiltonian Monte Carlo (HMC) via STAN for model fitting, with 2500 samples on 2 independent chains with a burn-in of 500 \citep{stan_development_team_stan_2024}.  An advantage of Hamiltonian Monte Carlo is samples are nearly independent so one more quickly achieves a sampling error that is an order of magnitude smaller than estimation error.  
\begin{figure}[H]
\begin{subfigure}[h]{0.48\linewidth}
\includegraphics[width=1.0\textwidth]{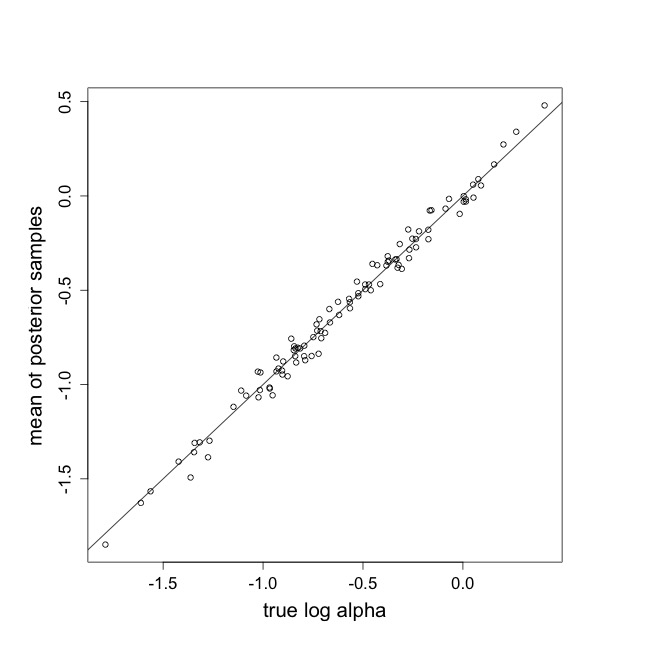} 
\caption{Small variance simulation for $\alpha$ on 3 followups.}
\end{subfigure}
\hspace{0.2cm}
\begin{subfigure}[h]{0.48\linewidth}
\includegraphics[width=1.0\textwidth]{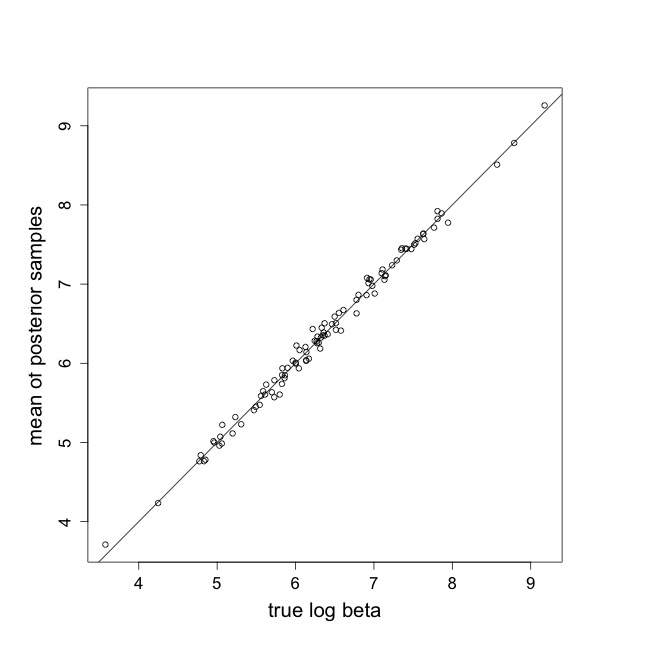} 
\caption{Small variance simulation for $\beta$ on 3 followups.}
\end{subfigure}
\caption{True parameters versus means of posterior samples, 3 followups. 45 degree line indicates perfect agreement.}
\label{fig:logest}
\end{figure}
\begin{figure}[H]
\begin{subfigure}[h]{0.48\linewidth}
\includegraphics[width=1.0\textwidth]{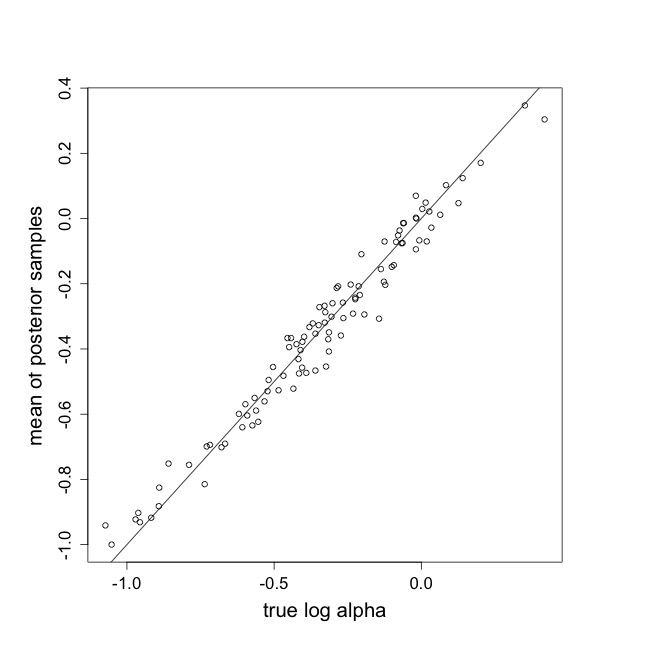} 
\caption{Big variance simulation for $\alpha$ on 2 followups.}
\end{subfigure}
\hspace{0.2cm}
\begin{subfigure}[h]{0.48\linewidth}
\includegraphics[width=1.0\textwidth]{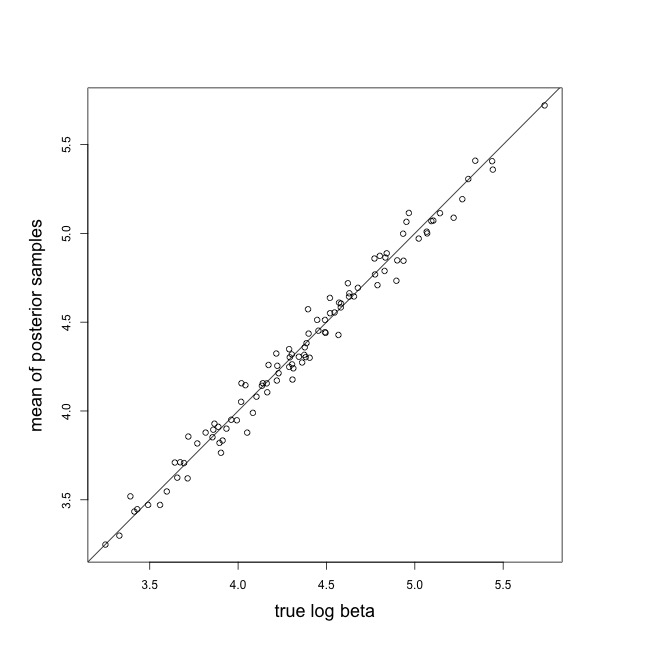} 
\caption{Big variance simulation for $\beta$ on 2 followups.}
\end{subfigure}
\caption{True parameters versus means of posterior samples, 2 followups. 45 degree line indicates perfect agreement.}
\label{fig:logest2}
\end{figure}
\begin{figure}[H]
\begin{subfigure}[h]{0.52\linewidth}
\includegraphics[width=1.0\textwidth]{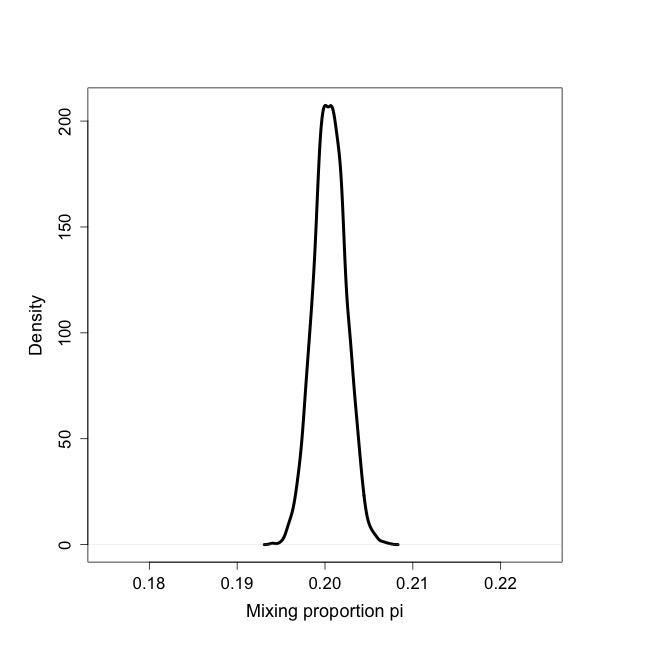} 
\caption{Distribution of samples of $\pi$ for simulation with larger variance parameterization and on 2 followups.}
\end{subfigure}
\hspace{0.2cm}
\begin{subfigure}[h]{0.52\linewidth}
\includegraphics[width=1.0\textwidth]{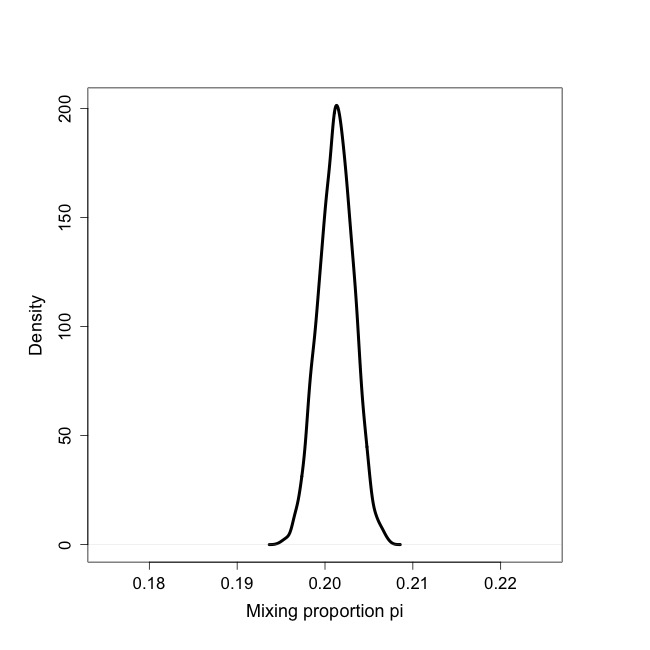} 
\caption{Distribution of samples of $\pi$ for simulation with smaller variance parameterization and on 3 followups.}
\end{subfigure}
\caption{Posterior $\pi$ samples under different simulation scenarios.}
\label{fig:mix_pi}
\end{figure}
Simulations indicate the fitting procedure characterizes the parameters used in the data generating mechanism well, with Figures \ref{fig:logest} and \ref{fig:logest2} suggesting tight alignment between true $\alpha_i$ and $\beta_i$ and the mean of posterior samples.  Likewise, Figure \ref{fig:mix_pi} suggests samples of logit $\pi$ align closely under inverse logit transformation to the true parameter of 0.2, with the distribution nearly perfectly centered at the value. The small degree of bias is likely induced by fitting the parameter in symmetric logit space prior to application of the inverse function.  
\begin{figure}[H]
\begin{subfigure}[h]{0.52\linewidth}
\includegraphics[width=1.0\textwidth]{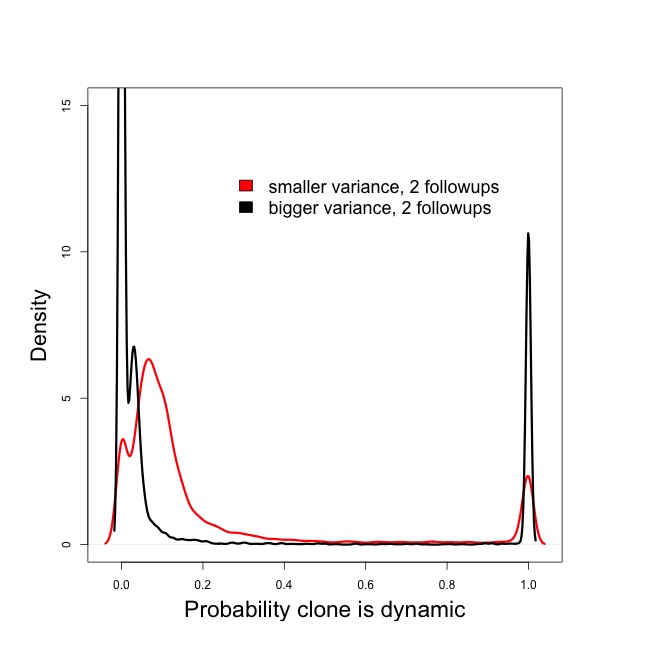} 
\caption{Component membership of clones under 2 followups, bigger and smaller variance.}
\end{subfigure}
\hspace{0.2cm}
\begin{subfigure}[h]{0.52\linewidth}
\includegraphics[width=1.0\textwidth]{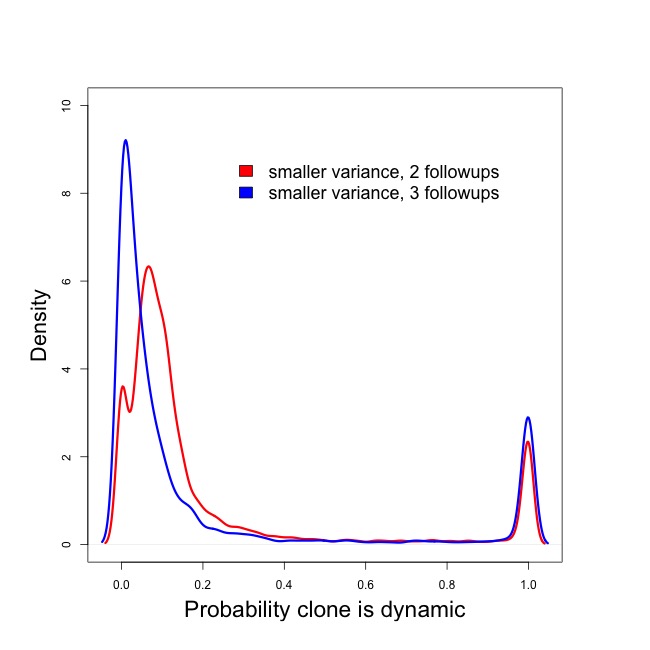} 
\caption{Component membership of clones under smaller variance, 2 and 3 followups.}
\end{subfigure}
\caption{Density plots of mean probability component membership under different simulation scenarios.}
\label{fig:mix}
\end{figure}
\begin{figure}[H]
\centering
\includegraphics[width=0.65\textwidth]{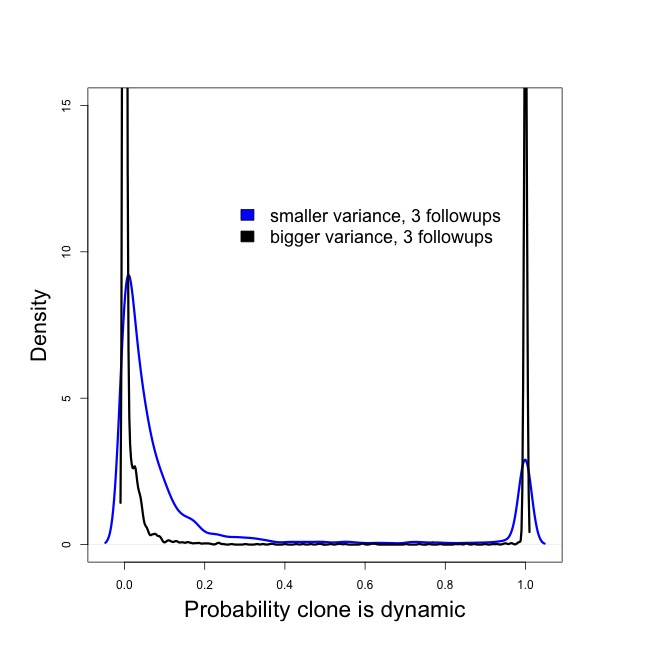} 
\caption{Densities of component membership of clones under 3 followups, smaller and bigger variance.}
\label{fig:mix_second}
\end{figure}
Figures \ref{fig:mix} and \ref{fig:mix_second} likewise show the model discriminating well between static and dynamic clones as intended.  The left hand mode in each density plot corresponds to static clones, while the right hand mode corresponds to dynamic.  The peakedness of the modes varies by plot as expected--where there is more information to discriminate dynamic versus static behavior with 3 followups rather than 2, the peakedness is greater.  Likewise, in the simulation scenarios using a Gamma prior with on average greater variation for the subsequent Poisson mean parameter, the peakedness is greater because dynamic behavior is easier to identify, other things equal.

\section{Derivation of static and dynamic mixture components}
\label{sec:derive}

\subsection{Deriving the static model}

Since for the static component we have

$$C_{ijk} \,|\, (\lambda_{ij}, O_{jk}, D_{ij}=0) \sim \mbox{Pois}(\lambda_{ij} O_{jk} )$$

\noindent
and

$$\lambda_{ij} | (D_{ij}=0) \sim \mbox{Gamma}(\alpha_j, \beta_j),$$

\noindent
then the probability mass function (pmf) for the likelihood and density for the prior are

$$l(C_{ij\cdot} | \lambda_{ij} ,  O_{j\cdot }) = \prod_{k=1}^{T_j} \frac{\exp(-\lambda_{ij}O_{jk})\, (\lambda_{ij} O_{jk})^{C_{ijk}}}{C_{ijk}!} = \frac{\exp(- \lambda_{ij} \sum_k O_{jk}) \lambda_{ij}^{\sum_k C_{ijk}} \prod_{k=1}^{T_j} O_{jk}^{C_{ijk}}}{\prod_{k=1}^{T_j} C_{ijk}!}$$

$$p(\lambda_{ij} | D_{ij} = 0, \alpha_j, \beta_j) =  \frac{\beta_j^{\alpha_j}}{\Gamma(\alpha_j)} \lambda_{ij}^{\alpha_j-1} \exp{(-\beta_j \lambda_{ij})},$$
respectively.  Taking the product of these terms and integrating out $\lambda_{ij}$ yields

\begin{align}
p(C_{ij\cdot } |O_{j\cdot}, D_{ij}=0, \alpha_j, \beta_j)=&\int p(C_{ij\cdot },\lambda_{ij} | O_{j\cdot }, D_{ij}=0, \alpha_j, \beta_j) \, d\lambda_{ij} \nonumber = \int l(C_{ij\cdot} | \lambda_{ij}, \, O_{j\cdot})p(\lambda_{ij} | D_{ij} , \alpha_j, \beta_j) \\
=& \frac{\beta_j^{\alpha_j} \,  \prod_{k=1}^{T_j} O_{jk}^{C_{ijk}}}{ \Gamma(\alpha_j) \, \prod_{k=1}^{T_j} C_{ijk}! }  \int_0^\infty  \lambda_{ij}^{\sum_k C_{ijk} + \alpha_j-1} \exp{(-\lambda_{ij}(\beta_j+ \sum_k O_{jk}))} d\lambda_{ij} \nonumber \\ 
 =& \frac{\beta_j^{\alpha_j} \,  \prod_{k=1}^{T_j} O_{jk}^{C_{ijk}}}{ \Gamma(\alpha_j) \, \prod_{k=1}^{T_j} C_{ijk}!} \cdot \frac{ \Gamma(\sum_k C_{ijk} + \alpha_j)}{(\beta_j + \sum_k O_{jk})^{\sum_k C_{ijk} + \alpha_j}}  \nonumber \\ 
=&\frac{ \Gamma(\sum_k C_{ijk} + \alpha_j) }{ \Gamma(\alpha_j) \, \prod_{k=1}^{T_j} C_{ijk}!}  \biggl(\frac{ \beta_j}{\beta_j + \sum_k O_{jk}}\biggr)^{\alpha_j}  \prod_{k=1}^{T_j} \biggl( \frac{O_{jk}}{\beta_j + \sum_k O_{jk}}\biggr)^{C_{ijk} }
\label{eqn:second}
\end{align}

\noindent
where we have used conjugacy to evaluate the integral, recognizing the kernel of the Gamma on updated parameters $\alpha_j'= \sum_k C_{ijk} + \alpha_j $ and $\beta_j'= \sum_k O_{jk} + \beta_j$.  The integral with respect to $\lambda_{ij}$ yields the inverse normalizing constant of the Gamma($\alpha_j',\beta_j'$).

\subsection{Deriving the dynamic model}

For the dynamic component clonal likelihood we have

$$C_{ijk} \,|\, (\lambda_{ijk}, O_{jk}, D_{ij}=1) \sim \mbox{Pois}(\lambda_{ijk} O_{jk} )$$

\noindent
and for the prior

$$\lambda_{ijk} | (D_{ij}=1) \sim \mbox{Gamma}(\alpha_j, \beta_j)$$

\noindent
so that for each $k$ the pmf and density expressions are 
$$l(C_{ijk} | \lambda_{ijk}, \, O_{j\cdot}) = \frac{\exp(- \lambda_{ijk} \, O_{jk}) \lambda_{ijk}^{ C_{ijk}}  O_{jk}^{C_{ijk}} }{ C_{ijk}!} $$

$$p(\lambda_{ijk}  | D_{ij} = 1, \alpha_j, \beta_j) = \frac{\beta_j^{\alpha_j}}{\Gamma(\alpha_j)} \lambda_{ijk}^{\alpha_j-1} \exp{(-\beta_j \lambda_{ijk})}, $$
respectively.  The product across $k$ then is
\begin{align}
p(C_{ij\cdot}\,  |\, O_{j\cdot} ,D_{ij}=1, & \alpha_j, \beta_j) =  \prod_{k=1}^{T_j} \int l(C_{ijk} | \lambda_{ijk}, \, O_{j\cdot})p(\lambda_{ijk}  | D_{ij} = 1, \alpha_j, \beta_j) \, d\lambda_{ijk} \nonumber \\ 
= &  \prod_{k=1}^{T_j} \int \frac{\exp(- \lambda_{ijk} \, O_{jk}) \lambda_{ijk}^{ C_{ijk}}  O_{jk}^{C_{ijk}} }{ C_{ijk}!} \cdot \frac{\beta_j^{\alpha_j}}{\Gamma(\alpha_j)} \lambda_{ijk}^{\alpha_j-1} \exp{(-\beta_j \lambda_{ijk})} 
\,  d\lambda_{ijk}  \nonumber \\
=  & \int \frac{\exp(- \sum_k \lambda_{ijk} ( O_{jk} + \beta_j) ) \prod_{k=1}^{T_j} O_{jk}^{C_{ijk}} \, \lambda_{ijk}^{ C_{ijk}+\alpha_j-1}  }{\prod_{k=1}^{T_j} C_{ijk}!} \cdot \frac{\beta_j^{\alpha_j}}{\Gamma(\alpha_j)}  \,  d\lambda_{ijk}  \nonumber \\
%
= & \prod_{k=1}^{T_j} \frac{ \Gamma( C_{ijk} + \alpha_j) }{ \Gamma(\alpha_j)  C_{ijk}!}  \biggl(\frac{ \beta_j}{\beta_j + O_{jk}}\biggr)^{\alpha_j}   \biggl( \frac{O_{jk}}{\beta_j + O_{jk}}\biggr)^{ C_{ijk}} \label{eqn:dyn}
\end{align}
where again we have used conjugacy to evaluate the integral, recognizing the kernel of the Gamma on updated parameters $\alpha_j'= \sum_k C_{ijk} + \alpha_j $ and $\beta_j'= \sum_k O_{jk} + \beta_j$.  The integral with respect to $\lambda_{ijk}$ yields the inverse normalizing constant of the Gamma($\alpha_j',\beta_j'$).

\section{HLA typing and clonal expansions}
\label{sec:hla}
HLA typing and baseline-followup data were available for analysis on 23 patients. Results showed modest associations between presence of the A*03 and C*07 alleles and clonal expansions.  
\begin{figure}[H]
\begin{subfigure}[h]{0.52\linewidth}
\includegraphics[width=1.0\textwidth]{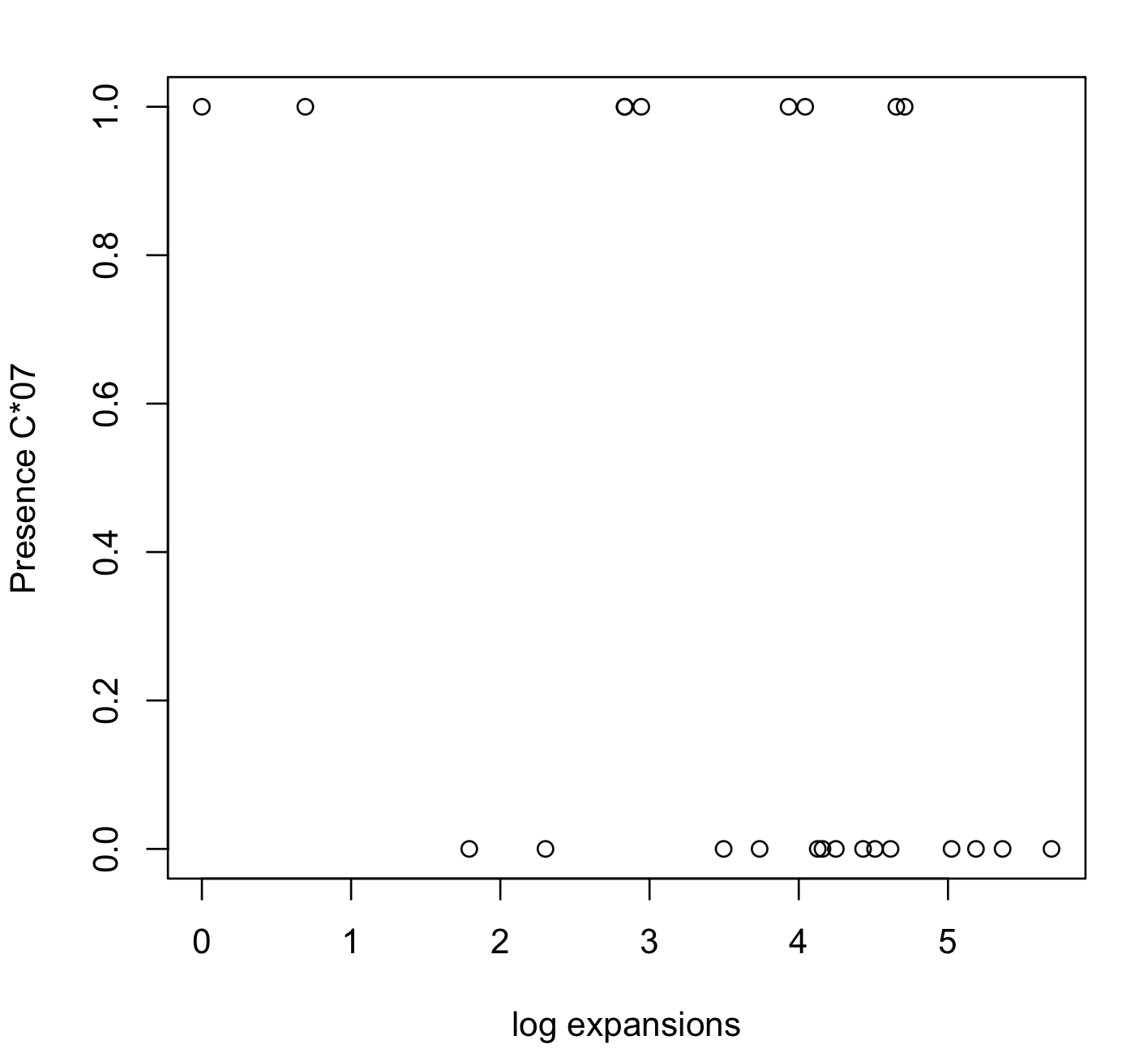} 
\caption{P-value of linear model under log transform, 0.043}
\end{subfigure}
\hspace{0.2cm}
\begin{subfigure}[h]{0.52\linewidth}
\includegraphics[width=1.0\textwidth]{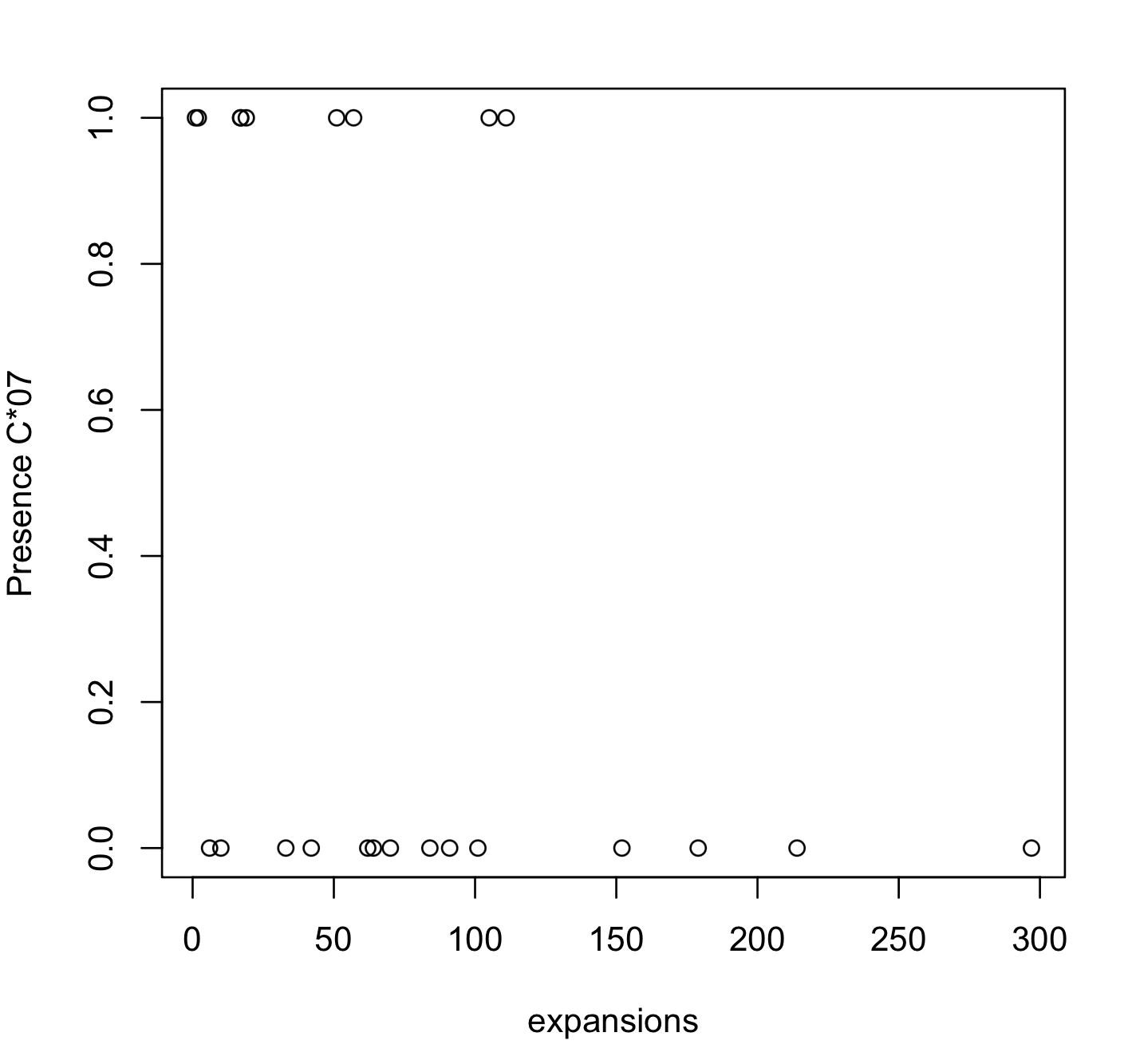} 
\caption{Visual depiction of untransformed association.}
\end{subfigure}
\vspace{2cm}
\phantom{ASDFASDFASDFASDF}
\begin{subfigure}[h]{0.52\linewidth}
\includegraphics[width=0.99\textwidth]{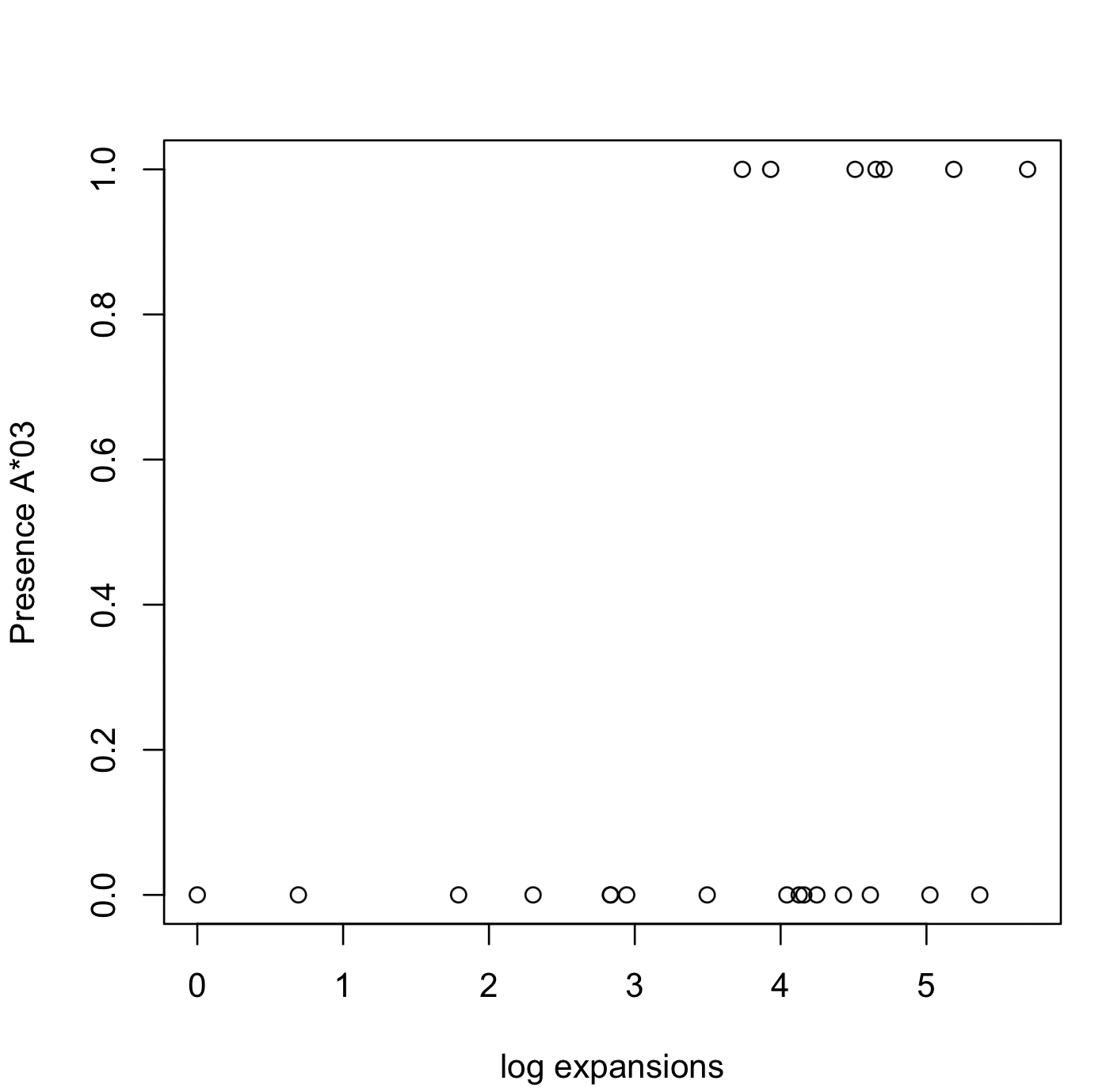} 
\caption{P-value of linear model under log transform, 0.040}
\end{subfigure}
\label{fig:mix2}
\end{figure}

\end{document}